\documentclass[preprin,showpacs,preprintnumbers,eqsecnum,amsmath,amssymb]{revtex4}
\usepackage{dcolumn}
\usepackage[dvips]{epsfig}

\usepackage{amsmath}
\usepackage{changes}
\usepackage{ulem}
\usepackage[thicklines]{cancel}

\begin{document}
\baselineskip=0.8 cm
\title{\bf Quantum-classical correspondence in resonant and nonresonant Rabi-Stark model}

\author{Shangyun Wang$^{1}$\footnote{Corresponding author: sywang@hynu.edu.cn}, Songbai Chen$^{2,3}$\footnote{csb3752@hunnu.edu.cn}, Jiliang
Jing$^{2,3}$ \footnote{jljing@hunnu.edu.cn}}

\affiliation{$^{\textit{1}}$College of Physics and Electronic Engineering, Hengyang Normal University, Hengyang 421002, China
 \\ $^{\textit{2}}$Department of Physics, Key Laboratory of Low Dimensional Quantum Structures
and Quantum Control of Ministry of Education, Synergetic Innovation Center for Quantum Effects and Applications, Hunan
Normal University,  Changsha, Hunan 410081, People's Republic of China
\\
$^{\textit{3}}$Center for Gravitation and Cosmology, College of Physical Science and Technology, Yangzhou University, Yangzhou 225009, People's Republic of China}

\begin{abstract}
\baselineskip=0.6 cm
\begin{center}
{\bf Abstract}
\end{center}

Testing the correspondence principle in nonlinear quantum systems is a fundamental pursuit in quantum physics.
In this paper, we employed mean field approximation theory to study the semiclassical dynamics in the Rabi-Stark model (RSM) and showed that the nonlinear Stark coupling significantly modulates the semiclassical phase space structure.
By analyzing the linear entanglement entropy of coherent states prepared in the classical chaotic and regular regions of the semiclassical phase space, we demonstrate that quantum-classical correspondence can be achieved in the RSM with large atom-light frequency ratios.
While this correspondence fails in the resonant Rabi model because its truncated photon number is insufficient to approach the large quantum number limit, we discovered that in the resonant RSM when the nonlinear Stark coupling $U\rightarrow\pm1$, the time-averaged linear entanglement entropy correlates strongly with the semiclassical phase space.
In particular, when $U \rightarrow -1$, the truncated photon number in the resonant RSM is very close to that in the resonant Rabi model, but the time-averaged linear entanglement entropy still corresponds well with the semiclassical phase space.
This result demonstrates that quantum-classical correspondence can be realized in the few-body resonant RSM.

\end{abstract}
\maketitle

\newpage
\section{Introduction}

The Correspondence Principle, as proposed by the Copenhagen school of quantum mechanics, states that the dynamics of a quantum system should converge to classical dynamics in the large quantum number limit. In classical mechanics, the state of a physical system is generally characterized by momentum and coordinate in phase space. Periodic orbits in phase space correspond to regular motion, and chaotic orbits are aperiodic trajectories which exhibit extreme sensitivity to initial conditions. In quantum mechanics, characterizing this sensitivity in chaotic systems is extremely challenging because the correspondence principle does not allow conjugate observables such as position and momentum to take on accurate values at the same time. This fundamental incompatibility poses a challenge to quantum-classical correspondence\cite{qcc} and has motivated a long-standing search for quantum tools which exhibit significant differences between chaotic and regular regions in semiclassical chaotic systems\cite{csqs1,csqs2,csqs3,csqs4,csqs5}.

Since chaos is inherently a dynamical phenomenon, special significance attaches to quantum dynamical signatures such as the generation of entanglement entropy\cite{gee1,gee2,gee3,gee4}, the Husimi quasiprobability distribution function\cite{Husimi1,Husimi2,Husimi3,Husimi4,Husimi5,Husimi6,Husimi7,Husimi8} and the Loschmidt Echo which exhibits sensitivity to perturbation\cite{leo1,leo2,leo3,leo4,leo5}.
In particular, the significant difference in linear entanglement entropy between chaotic and regular regions has been experimentally observed in multi-body quantum systems\cite{Husimi2}.
The correspondence between entanglement dynamics and phase space serves as a test of the classical-quantum correspondence in quantum chaotic systems, including the quantum kicked top model\cite{ktp1,ktp2,ktp3,ktp4,ktp5,ktp6,ktp7}, the spin chains system \cite{spinc1,spinc2,spinc3,spinc4}, and the Dicke model\cite{gee2,dicke1,dicke2,dicke3,dicke4,dicke5,dicke6}.
The inquiry into the connection between quantum chaos and entanglement has recently been expanded to the quantum Rabi model which describes the fundamental interaction between a two-level atom and a single-mode bosonic field\cite{chaos in rabi1,chaos in rabi2}.
Within this model, the quantum-classical correspondence is governed by the truncated photon number and is manifested under conditions of large atom-field frequency ratios\cite{chaos in rabi3}. However, this physical condition is inherently contradictory to resonance.

On the other hand, the resonance ensures a precise match between the driving frequency and the transition frequency of an atom. This enables the observation of the most distinct and stable Rabi oscillations in experiments\cite{Res Rabi1, Res Rabi2}, as well as the achievement of the phase shift between two quantum bits---a key step toward constructing quantum logic gates\cite{Res Rabi3,Res Rabi4}.
Moreover, the resonance interaction between atom and cavity is crucial for achieving high-fidelity and high-efficiency coherent exchange of quantum states\cite{Res Rabi5}.
It ensures reliable transfer of quantum states between quantum nodes and transmission channels, thereby laying the foundation for realizing entanglement distribution and quantum state transfer among distant quantum nodes\cite{Res Rabi6}.
These milestone works demonstrate that the resonance effect in the Rabi model is not merely a special parameter, but rather akin to a master key which unlocks the strong-coupling quantum physics, precise quantum manipulation and quantum communication. In the semiclassical limit, the resonant single-atom-cavity system may also exhibit some novel physical phenomena. Being far from the thermodynamic limit, the resonant Rabi model lacks a solid theoretical foundation for studying its semiclassical dynamics and the corresponding quantum features. The fundamental contradiction between resonance effects and the semiclassical limit in the Rabi model forces us to introduce a new action term in the model as a feasible method for exploring its semi classical dynamics.

Recently, Cong et al. engineered Stark interactions through laser-driven transitions and successfully simulated the dynamical behavior of the Rabi-Stark model in a single trapped-ion system\cite{cong}. This breakthrough in quantum simulation has sparked widespread interest in the Rabi-Stark model.
The Rabi-Stark model, comprising the Rabi model with an additional nonlinear Stark coupling term, was first proposed by Grimsmo and Parkins\cite{RSM1,RSM2}.
Due to the Stark coupling, the energy spectrum of RSM exhibits some interesting features such as first-order phase transition and spectral collapse\cite{RSMps1,RSMps2,RSMps3}. The first-order and continuous quantum phase transitions critically modulate the efficiency and power of anisotropic quantum RSM-based engines\cite{RSMps4}.
In single-atom-cavity systems, the interplay between quantum interference and the strong vacuum Rabi splitting which is induced by the Stark shift results in strong photon blockade\cite{RSMps5,RSMps6,RSMps7}.
Furthermore, the Stark coupling directly manipulates the photon squeezing effects in RSM, enabling substantial enhancement or suppression\cite{RSMps8}. In particular, Ref.\cite{RSMps9}proved that if the Stark coupling strength is the same as the cavity frequency, the Stark coupling strength $U \rightarrow \pm1$ corresponds to the thermodynamic limit in the RSM. The definition of this thermodynamic limit is distinct from those in both the quantum Rabi model and the Dicke model: the former corresponds to the atom-field frequency ratio approaching infinity\cite{limit1}, while the latter corresponds to the number of atoms approaching infinity\cite{limit2}. Therefore, it provides us with a possible perspective for studying the semiclassical dynamics and quantum chaos in the resonant RSM.

In this paper, we utilize the correspondence between the semiclassical phase space and the time-averaged entanglement entropy to investigate the quantum-classical correspondence in both resonant and nonresonant RSM.
This paper is structured as follows: In Sec. \ref{sec2}, we briefly introduce the quantum and semiclassical RSM. In Sec. \ref{sec3}, we investigate the impact of Stark coupling on the phase-space structure of the QRM at large atom-field frequency ratios, and calculate the time-averaged entanglement entropy over the entire phase space. In Sec. \ref{sec4}, we discuss the quantum-classical correspondence in the resonant QRM. Finally, we
present results and a brief summary.

\section{Quantum and Classical Rabi-Stark Model}\label{sec2}
\begin{figure}[ht]
\begin{center}
\includegraphics[width=16cm]{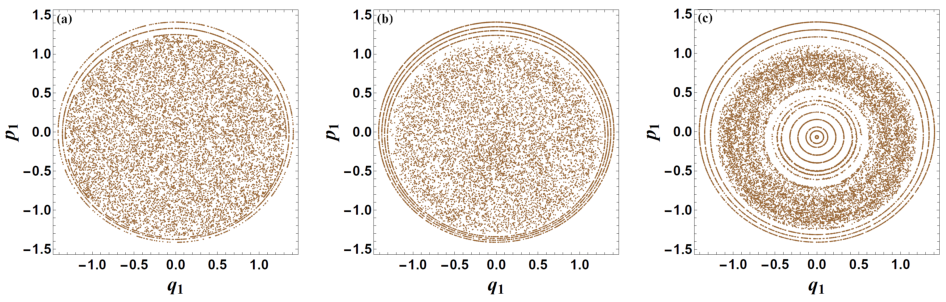}
\caption{The Poincar\'{e} sections for the RSM in the case: $q_{2}=0, p_{2}>0$, with $\omega=14$, $\omega_{0} =1$, $g=5$ and the system energy $E = 7$.  The Stark coupling parameters from left to right are U = -0.3, 0, and 0.3, respectively.}\label{NonRSM-ps}
\end{center}
\end{figure}
\begin{figure}[ht]
\begin{center}
\includegraphics[width=16cm]{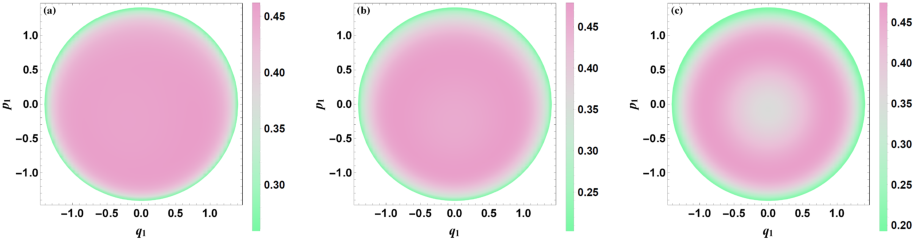}
\caption{The distribution of the time-averaged linear entanglement entropy of Poincar\'{e} sections in Fig. \ref{NonRSM-ps}. The Stark coupling parameters from left to right are U = -0.3, 0, and 0.3, respectively. Here, the time interval is $t\in[0, 500]$ and the system photon number is truncated at $180$, $160$ and $150$, respectively.}\label{NonRSM-MEE}
\end{center}
\end{figure}
Let us now briefly introduce the quantum Rabi-Stark model which is one of the simplest and
most fundamental models describing quantum light-matter interaction with Stark coupling.
The quantum Rabi-Stark Hamiltonian can be expressed as ($\hbar=1$)
\begin{eqnarray}
\hat{H}&=& (\frac{\omega}{2} + U\hat{a}^{\dagger}\hat{a} )\hat{\sigma}_{z} + \omega_0 \hat{a}^{\dagger}\hat{a} + g(\hat{a}^{\dagger} + \hat{a})(\hat{\sigma}_{+}+ \hat{\sigma}_{-}), \label{ha1}
\end{eqnarray}
where $\omega$ is the frequency of a two-level system, and $\hat{\sigma}_{z}$ is the usual Pauli matrix describing the two-level system. $\hat{a}^{\dagger}$ ($\hat{a}$) are the creation (annihilation) bosonic operators of the cavity mode with frequency $\omega_0$. The coupling $g$ denotes the linear coupling strength between the qubit and the cavity, and
$\hat{\sigma}_{+}$, $\hat{\sigma}_{-}$ are atomic raising and lowering operators, respectively.
U is the Stark coupling strength and is determined by the dispersive energy shift, and when $U=0$, the Hamiltonian \ref{ha1} reduces to the quantum Rabi model.

To investigate the relationship between the linear entanglement entropy and the semiclassical phase space in the RSM, we choose coherent states as the initial states since they represent minimum-uncertainty wave packets in the classical phase space. As in Refs.\cite{gee2, Husimi2, ctp}, we take the initial quantum states as
\begin{eqnarray}
|\psi(0)\rangle &=& |\tau\rangle\otimes|\beta\rangle,\label{init1}
\end{eqnarray}
with
\begin{eqnarray}
|\tau\rangle &=& (1+\tau\tau^{*})^{-\frac{1}{2}}e^{\hat{\tau} \sigma_{+}}|\frac{1}{2},-\frac{1}{2}\rangle,\label{acoh1}\\
|\beta\rangle &=& e^{-\beta\beta^{*}/2}e^{\beta \hat{a}^{\dagger}}|0\rangle, \label{bcoh1}
\end{eqnarray}
and
\begin{eqnarray}
\tau &=& \frac{q_1 + i p_1}{\sqrt{2-q^{2}_1 -p^{2}_1}},  \ \ \ \ \ \beta = (q_2 + ip_2)/\sqrt{2},   \label{tb1}
\end{eqnarray}
where $|\tau\rangle$ and $|\beta\rangle$ are the Bloch coherent states of atom and the Glauber coherent states of bosons. The states $|\frac{1}{2},-\frac{1}{2}\rangle$ and $|0\rangle$ denote the ground state of a two-level atom and the vacuum state of the single-mode cavity field, respectively.
The variables $q_1, p_1, q_2, p_2$ describe the phase space of the system, and the indices $1$ and $2$ denote the atomic and single-mode boson field subsystems, respectively.
Using the mean-field approximation and the relations
\begin{eqnarray}
\langle\tau|\hat{\sigma}_{+}|\tau\rangle &=& \frac{\tau^*}{1+\tau\tau^*},
\end{eqnarray}
\begin{eqnarray}
\langle\tau|\hat{\sigma}_{-}|\tau\rangle &=& \frac{\tau}{1+\tau\tau^*},
\end{eqnarray}
\begin{eqnarray}
\langle\tau|\hat{\sigma}_{z}|\tau\rangle &=& -\frac{1-\tau\tau^*}{1+\tau\tau^*},
\end{eqnarray}
\begin{eqnarray}
\langle\beta|\hat{a}|\beta\rangle &=& \beta,
\end{eqnarray}
\begin{eqnarray}
\langle\beta|\hat{a}^{\dagger}|\beta\rangle &=& \beta^*,
\end{eqnarray}
one can obtain the semiclassical Hamiltonian related to the quantum RSM,
\begin{eqnarray}
H_{cl} &\equiv& \langle \tau\beta|\hat{H}|\tau\beta\rangle = \frac{1}{2} ((q^{2}_1 +p^{2}_1 -1)((q^{2}_2 + p^{2}_2)U + \omega)+ (q^{2}_2 + p^{2}_2)\omega_0) + g q_1 q_2\sqrt{4-2(q^{2}_1 +p^{2}_1)},\label{ha2}
\end{eqnarray}
where $q_2=(a^{\dag} + a)/\sqrt{2}$ and $p_2= i(a^{\dag} - a)/\sqrt{2}$. Then, the corresponding Hamilton's equations of motion are given by

\begin{eqnarray}
\dot{q_1} &=&  p_1 (\omega +U(q^{2}_2 +p^{2}_2)) - \frac{2 g q_1 p_1 q_2}{\sqrt{4- 2(q^{2}_1 +p^{2}_1)}} ,\label{deq1}
\end{eqnarray}
\begin{eqnarray}
\dot{p_1} &=& -q_1 (\omega +U(q^{2}_2 +p^{2}_2)) + \frac{\sqrt{2} g q_2 (2q^{2}_1+ p^{2}_1 -2 )}{\sqrt{2- q^{2}_1 -p^{2}_1)}} ,\label{deq2}
\end{eqnarray}
\begin{eqnarray}
\dot{q_2} &=& p_2 (\omega_0 +U(q^{2}_1 +p^{2}_1 - 1)),\label{deq3}
\end{eqnarray}
\begin{eqnarray}
\dot{p_2} &=& -q_2 \omega_0 - q_2 U(q^{2}_1 + p^{2}_1 - 1)-g q_1 \sqrt{4- 2(q^{2}_1 +p^{2}_1)}.\label{deq4}
\end{eqnarray}
\begin{figure}[ht]
\begin{center}
\includegraphics[width=16cm]{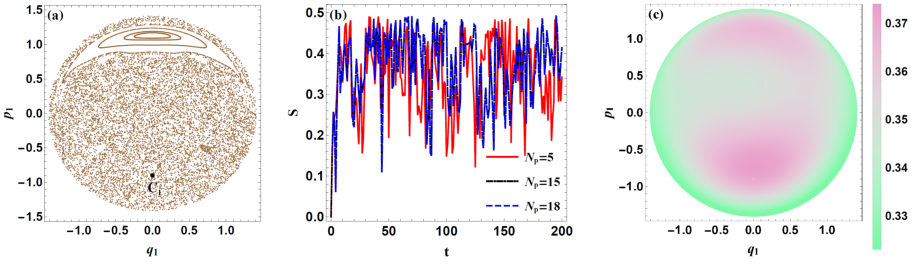}
\caption{The correspondence between classical dynamics and quantum linear entanglement entropy for the resonant Rabi model with system energy $E = 1.3$. Part (a) is the classical Poincar\'{e} section in the case: $q_{2}=0, p_{2}>0$, with $\omega=\omega_{0} =1$, $g=0.4$ and the coordinates of point $C_1$ are $q_{1} = 0,p_{1} =-0.9,q_{2} =0,p_{2} = 1.67033$. Part (b) shows the time evolution of the linear entanglement entropy $S(t)$ with different initial system photon numbers $N_p$ for the initial state centred at point $C_{1}$. Part (c) shows the distribution of the time-averaged entanglement entropy corresponding to the Poincar\'{e} section in part (a). Here, the photon number was truncated at 18. }\label{RRM}
\end{center}
\end{figure}
\section{Quantum-classical Correspondence of Rabi-Stark Model with Large Atom-Field
frequency ratios }\label{sec3}

Recently, Ref. \cite{RSMps5} showed that the negative (positive) Stark shift results in a significant increase in the lower (higher) branch of the dressed-state and a decrease in the higher (lower) branch.
Moreover, Wang et al. showed that the critical coupling strength of the superradiant phase transition in the Dicke-Stark model decreases with a positive Stark shift and increases with a negative one\cite{dickestark}. This suggests that Stark couplings may exhibit different effects on the dynamics in the RSM. In this section, we study the semiclassical phase space structure of the Rabi model with different Stark coupling strengths, and investigate the distribution of quantum entanglement in the semiclassical phase space.

With the classical dynamic equations (Eqs.\ref{deq1}-\ref{deq4}), we obtain the Poincar\'{e} section of the RSM with large atom-light frequency ratios, as shown in Fig. \ref{NonRSM-ps}.
These mixed phase space sections contain both stable islands and chaotic seas, where the stable islands are represented by rings composed of many discrete points, while the chaotic seas are represented by surfaces formed by discrete points. Motion across the boundaries between
regular and chaotic regions is classically forbidden. Comparing Figs. \ref{NonRSM-ps}(a) and \ref{NonRSM-ps}(b), it is observed that a negative Stark coupling leads to an expansion of the chaotic region in the phase space of the RSM. In contrast, positive Stark coupling constant results in a reduction of the chaotic region, as shown when comparing Figs. \ref{NonRSM-ps}(b) and \ref{NonRSM-ps}(c). This demonstrates that the Stark coupling effectively modulates the semiclassical phase space structure of the RSM.
\begin{figure}[ht]
\begin{center}
\includegraphics[width=16cm]{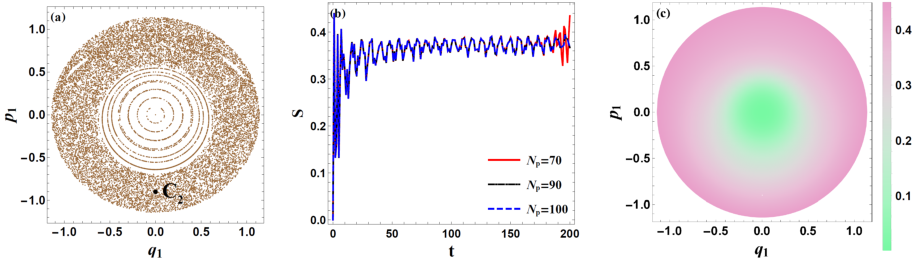}
\caption{The correspondence between classical dynamics and quantum linear entanglement entropy for the resonant RSM with system energy $E = 0.15$. Part (a) is the classical Poincar\'{e} section in the case: $q_{2}=0, p_{2}>0$, with $\omega=\omega_{0} =1$, $g=0.6$, $U=0.999$ and the coordinates of point $C_2$ are $q_{1} = 0,p_{1} =-0.9,q_{2} =0,p_{2} = 0.77769$. Part (b) is the time evolution of linear entanglement entropy $S(t)$ with different initial system photon numbers $N_p$ for the initial state centred at point $C_{2}$. Part (c) shows the distribution of time-averaged entanglement entropy corresponding to the Poincar\'{e} section in part (a). Here, the photon number was truncated at 100. }\label{RRSMUP1}
\end{center}
\end{figure}

The efficacy of linear entanglement entropy as a tool for investigating quantum chaotic behavior has been established through its extensive application in studies of systems such as the Dicke model\cite{gee2,dicke2}, the quantum kicked top\cite{Husimi2,ktp2} and the quantum Rabi model\cite{chaos in rabi2,chaos in rabi3}. Moreover, the distribution of the time-averaged linear entanglement entropy in the semiclassical Poicar\'{e} section serves as an effective tool for verifying the correspondence principle in quantum chaotic systems.
The time-averaged entanglement entropy
\begin{eqnarray}
S_{m}=\frac{1}{T}\int_{t_{1}}^{t_{2}}S(t)\rm{d}t,
\end{eqnarray}
with the linear entanglement entropy
\begin{eqnarray}
S(t) &=& 1- \rm{Tr}_{1} \rho_{1}(t)^2, \label{cha1}
\end{eqnarray}
and the reduced-density matrix
\begin{eqnarray}
\rho_{1} &=& \rm{Tr}_{2}|\psi(t)\rangle\langle\psi(t)|,   \label{cha2}
\end{eqnarray}
where $\rm{Tr}_{i}$ denotes the trace over the $i$-th subsystem($i=1,2$).
The quantity $S(t)$ describes the degree of purity of the subsystems.
In this paper, the truncation photon number is set to be identical to that in Refs. \cite{chaos in rabi2,chaos in rabi3}. It is slightly above the critical value, and beyond this critical value the entanglement dynamics become independent of the photon number.
In Fig. \ref{NonRSM-MEE}, we present the distribution of the time-averaged entanglement entropy on the Poincar\'{e} sections in Fig. \ref{NonRSM-ps}.
It is clearly observed that a significant dip in $S_{m}$ appears between the chaotic and regular regions, and initial states localized in chaotic seas exhibit more entanglement generation compared to those in the regular regions.
The correspondence between Fig. \ref{NonRSM-ps} and Fig. \ref{NonRSM-MEE} indicates that quantum-classical correspondence can still be achieved in the RSM with large atom-field frequency ratios.

\section{Quantum-classical Correspondence in the Resonant Rabi-Stark Model}\label{sec4}
\begin{figure}[ht]
\begin{center}
\includegraphics[width=16cm]{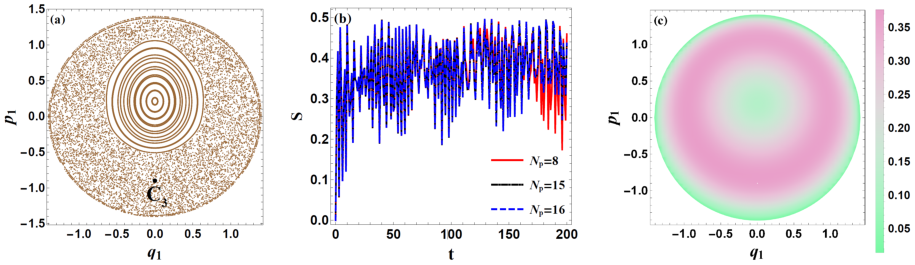}
\caption{The correspondence between classical dynamics and quantum linear entanglement entropy for the resonant RSM with system energy $E = 0.6$. Part (a) is the classical Poincar\'{e} section in the case: $q_{2}=0, p_{2}>0$, with $\omega=\omega_{0} =1$, $g=0.2$, $U=-0.999$ and the coordinates of point $C_3$ are $q_{1} = 0,p_{1} =-0.9,q_{2} =0,p_{2} = 1.08086$. Part (b) is the time evolution of linear entanglement entropy $S(t)$ with different the initial system photon number $N_p$ for the initial state centred at point $C_{3}$. Part (c) shows the distribution of time-averaged entanglement entropy corresponding to the Poincar\'{e} section in part (a).Here, the photon number was truncated at 16. }\label{RRSMUN1}
\end{center}
\end{figure}
As mentioned in the preceding section, achieving quantum-classical correspondence in the single-atom-cavity interaction system typically necessitates a large atom-field frequency ratio. This implies that the system is in a non-resonant state.
On the other hand, resonance conditions hold irreplaceable value in exploring quantum interaction mechanisms\cite{qim1,qim2,qim3}, quantum information processing\cite{qip1,qip2,qip3,qip4}, and discovering novel quantum phenomena\cite{dnqp1, dnqp2,dnqp3}. Therefore, introducing new interaction terms into the resonant Rabi model to achieve quantum-classical correspondence has become a non-negligible option.

In this section, we further explore the quantum-classical correspondence in the resonant Rabi model. The quantum-classical correspondence relationship in the resonant Rabi model is shown in Fig. \ref{RRM}. It can be observed that the entanglement entropy in the regular regions of the Poincar\'{e} section is even higher than that in the chaotic sea, as shown in Figs. \ref{RRM}(a) and \ref{RRM}(c). This indicates that the quantum-classical correspondence cannot be achieved in the resonant Rabi model. To elucidate this, we choose an initial point on the Poincar\'{e} section and trace the time evolution of the corresponding linear entanglement entropy with different initial photon numbers. In Fig. \ref{RRM}(b), we find that the time evolution of the entanglement entropy changes with increasing initial photon number. However, once the initial photon number exceeds a certain critical value, the entanglement entropy saturates and no longer changes. Therefore, the truncated photon number can be chosen to be slightly above this critical value.
The truncated photon number in the resonant Rabi model is significantly lower than that in the large atom-field frequency ratio regime. Thus, the resonant Rabi model should be viewed as a few-body system that is far from the thermodynamic limit. Consequently, this system exhibits a fundamental breakdown of quantum-classical correspondence.

Although quantum-classical correspondence cannot be achieved in the resonant Rabi model, recent studies demonstrate that the finite-frequency-ratio RSM system can approach the thermodynamic limit when the Stark coupling strength approaches the cavity frequency.
This provides a new approach to studying the quantum-classical correspondence in the resonant RSM. In Fig. \ref{RRSMUP1}, we present the quantum-classical correspondence in the resonant RSM in the limit of $U\rightarrow1$. We find that the entanglement entropy in the chaotic region of the Poincar\'{e} section is higher than that in the regular region, as shown in Figs. \ref{RRSMUP1}(a) and \ref{RRSMUP1}(c). This indicates that the quantum-classical correspondence can be achieved in the resonant RSM. In addition, we investigated the time evolution of entanglement entropy with different initial photon numbers, as shown in Fig. \ref{RRSMUP1}(b).
Comparing Fig. \ref{RRM}(b) and Fig. \ref{RRSMUP1}(b), we find that the entanglement dynamics in the resonant RSM with $U\rightarrow1$ exhibits a similar dependence on the initial photon number as in the resonant Rabi model, but its truncated photon number is about an order of magnitude higher.
Pinpointing the exact origin of the quantum-classical correspondence in the resonant RSM at $U\rightarrow1$ presents a persistent challenge. The ambiguity arises because the phenomenon could be attributed to two different thermodynamic limits: a large truncation photon number or $U\rightarrow1$.

When $U\rightarrow -1$, we find that chaotic regions are characterized by high entanglement entropy and regular regions exhibit low entanglement entropy, as shown in Figs. \ref{RRSMUN1}(a) and \ref{RRSMUN1}(c). This means that the quantum-classical correspondence in the resonant RSM can still be achieved. Nevertheless, we observe that the truncated photon number in the resonant RSM with $U \rightarrow -1$ is similar to that in the resonant Rabi model, as seen in Fig. \ref{RRSMUN1}(b) and Fig. \ref{RRM}(b). This indicates that $U\rightarrow -1$ can be regarded as a new thermodynamic limit and quantum-classical correspondence can be realized in the few-body resonant RSM.
It is noteworthy that the difference in the time-averaged entanglement entropy between chaotic and regular regions in the resonant RSM with $U\rightarrow -1$ can be observed experimentally.

\section{Conclusion}\label{sec5}

In this paper, we have investigated the semiclassical dynamics of the Rabi-Stark model using mean-field approximation theory, demonstrating that the nonlinear Stark coupling significantly modulates the semiclassical phase space structure.
We showed that quantum-classical correspondence can be achieved in the RSM with large atom-light frequency ratios, while it fails in the resonant Rabi model due to insufficient truncated photon numbers to approach the large quantum number limit.
Notably, in the resonant RSM, when the Stark coupling strength $U\rightarrow\pm1$, the time-averaged linear entanglement entropy strongly correlates with the semiclassical phase space.
In particular, when $U\rightarrow-1$, the truncated photon number is comparable to that in the resonant Rabi model, the quantum-classical correspondence remains valid.
These results indicate that quantum-classical correspondence can be realized in few-body resonant RSM systems, and offer a new perspective for exploring quantum chaos and entanglement dynamics in resonant light-matter interaction systems.

Our results open up new perspectives for the study of chaos and entanglement in few-body quantum systems. Future research could further explore the universal influence of different forms of nonlinear coupling on quantum-classical correspondence, develop quantum simulation experimental schemes based on this model, and validate its dynamical characteristics on practical quantum information platforms. Additionally, integrating open quantum system theory with non-equilibrium dynamical methods holds promise for revealing deeper connections among quantum chaos, decoherence, and information scrambling in resonant light-matter interaction systems, thereby advancing the extension and application of the quantum correspondence principle in complex coupled systems.

\section{\bf Acknowledgments}
This work was supported by the National Natural Science Foundation of China under Grant No.12275078,
11875026, 12035005, 2020YFC2201400, and the Scientifc Research Fund of Hunan Provincial Education Department under Grant No.24C0347. This work is also sponsored by the innovative research group of Hunan Province under Grant No. 2024JJ1006.

\end{document}